\documentclass{aa}

\usepackage{graphicx}
\usepackage{txfonts}

\usepackage{natbib} 
\bibpunct{(}{)}{;}{a}{}{,} 

\newcommand{\Mpc}{$h^{-1}$\thinspace Mpc}

\newcommand{\be}{\begin{equation}}
\newcommand{\ee}{\end{equation}}

\begin{document}

\title{Measuring galaxy segregation using the mark connection function}

\author{Vicent J. Mart\'{\i}nez\inst{1,2}
\and Pablo Arnalte-Mur\inst{1,2}
\and Dietrich Stoyan\inst{3}
}

\institute{Observatori Astron\`omic, Universitat de Val\`encia, Apartat
de Correus 22085, E-46071 Val\`encia, Spain
\and
Departament d'Astronomia i Astrof\'{\i}sica, Universitat de Val\`encia, E-46100-Burjassot, Val\`encia, Spain
\and
Institut f\"ur Stochastik, TU Bergakademie Freiberg, D-09596 Freiberg,
Germany
}


\authorrunning{V.J. Mart\'{\i}nez et al.}

\titlerunning{Mark connection function}

\offprints{V.J. Mart\'{\i}nez}

\abstract
{The clustering properties of galaxies belonging to different
luminosity ranges or having different morphological types are
different.  These characteristics or `marks' permit to understand
the galaxy catalogs that carry all this information as realizations
of marked point processes. Many attempts have been presented to
quantify the dependence of the clustering of galaxies on their inner
properties.}
{The present paper summarizes methods on spatial marked
statistics used in cosmology to disentangle luminosity, colour or
morphological segregation and introduces a new one in this context,
the mark connection function.}
{The methods used here are the partial correlation functions,
including the cross-correlation function, the normalised mark
correlation function, the mark variogram and the mark connection function. 
All these methods are applied to a
volume-limited sample drawn from the 2dFGRS, using the spectral type
$\eta$ as the mark.}
{We show the virtues of each method to provide information about the
clustering properties of each population, the dependence of the
clustering on the marks, the similarity of the marks as a function
of the pair distances, and the way to characterise the spatial
correlation between the marks.  We demonstrate by means of
these statistics that passive galaxies exhibit stronger spatial
correlation than active galaxies at small scales ($r \lesssim 20$
\Mpc ), and that the price galaxies have to pay to be close together is
having smaller values of the assigned marks, which means in our
case being more passive. Through the mark connection function we
quantify the relative positioning
of different types of galaxies within the overall clustering pattern.}
{The different marked statistics provide different information about 
the clustering properties of each population. Different aspects of 
the segregation are encapsulated by each measure, being the new one
introduced here --the mark connection function-- particularly useful 
for understanding the spatial correlation between the marks. }


\maketitle


\section{Introduction}

Galaxies of different morphological types show different clustering 
properties. It is well known, for example, that elliptical galaxies 
are preferentially found in  high density environments, such as the 
centres of rich galaxy clusters \citep{dressler1980}, while the dominant 
population of the field are mainly spiral galaxies 
\citep{davis1976,dressler1980}. Second order characteristics 
as the two point correlation function have been used to quantify the 
clustering of galaxies with different morphologies,
different spectral characteristics, different colours or belonging 
to different luminosity ranges 
\citep{phillipps1987,hamilton1988,davis1988,loveday1995,hermit1996, guzzo1997}. 
Bright galaxies show stronger spatial correlation 
than faint ones. Other clustering measures have been also used 
to quantify the luminosity or morphological
segregation: multifractals \citep{dominguez1989,  dominguez1994},
void probability functions \citep{vogeley1991, croton2004}, 
distributions of the distances to the nearest neighbours \citep{salzer1990}, etc.

The two-point correlation function $\xi(r)$ measures the excess 
probability of finding a neighbour at a distance $r$
from a given galaxy when compared with that probability for a homogeneous 
Poisson process. Morphological segregation
is encapsulated by the behaviour of $\xi(r)$ when it is calculated 
separately for different populations of galaxies. Elliptical galaxies 
show at small scales a correlation function with steeper slopes and larger 
amplitudes than  spirals \citep{loveday1995}. A recent analysis of the 
Two Degree Field Galaxy Redshift Survey (2dFGRS) has shown the same 
trend when comparing populations for different spectral types, being 
the two-point correlation function steeper for passive galaxies than 
for active galaxies \citep{mad03a}. Also, \citet{zehavi2002} 
have analysed the distribution of red and blue galaxies in the 
Sloan Digital Sky Survey (SDSS) by means of the projected correlation 
funcion $w_p(r_p)$ showing that red galaxies display a more prominent 
and steeper real-space correlation function than blue galaxies do.

The galaxy distribution can be considered a realisation of a point
process. However,  in many situations, each galaxy (point in the process)
carries additional information regarding a given characteristic
(e.g. morphological type) or a given numerical value that measures a
given galaxy property: luminosity, colour, spectral type. If we attach this
characteristic (mark) to the point in the process, we end up at a
marked point process, as it is called in mainstream spatial
statistics \citep{sto94a, martsaar02,ili08a}.

In this work, we compare different statistical methods for the study
of the marked galaxy distribution. We also introduce -- for the first
time in this context -- the mark connection function. We illustrate the usefulness
of these methods by applying them to a volume-limited sample drawn from the 2dFGRS 
with marks given by the galaxy spectral type. In Section 2, we describe
the sample and the marks assigned to the galaxies. In Section 3 we 
describe the different statistical methods considered, and in Section 4 we show
the results of applying them to our galaxy sample. In the conclusions, we
stress the capabilities of the mark connection function to 
characterise the spatial correlation between the marks.

\section{The 2dFGRS subsamples}

To illustrate the different mark clustering measures,
we used a nearly volume-limited sample drawn from the 2dFGRS
and prepared by the 2dF team 
\citep{croton2004}. It contains galaxies with absolute magnitudes 
in the range $-20 < M_{b_J} < -19$ at redshifts $z<0.13$.
In order to avoid the  effects of complicated boundaries while using a 
simple estimator, we selected galaxies inside a rectangular parallelepiped 
inscribed in the North slice of 2dFGRS. The final sample used contains 
$N=7741$ galaxies and covers a volume
of $V \sim 10^6$ (\Mpc)$^3$ where $h$ is the Hubble constant in units 
of 100~km~s$^{-1}$~Mpc$^{-1}$ .

\begin{figure*}[ht]
\centering
\includegraphics[width=17cm]{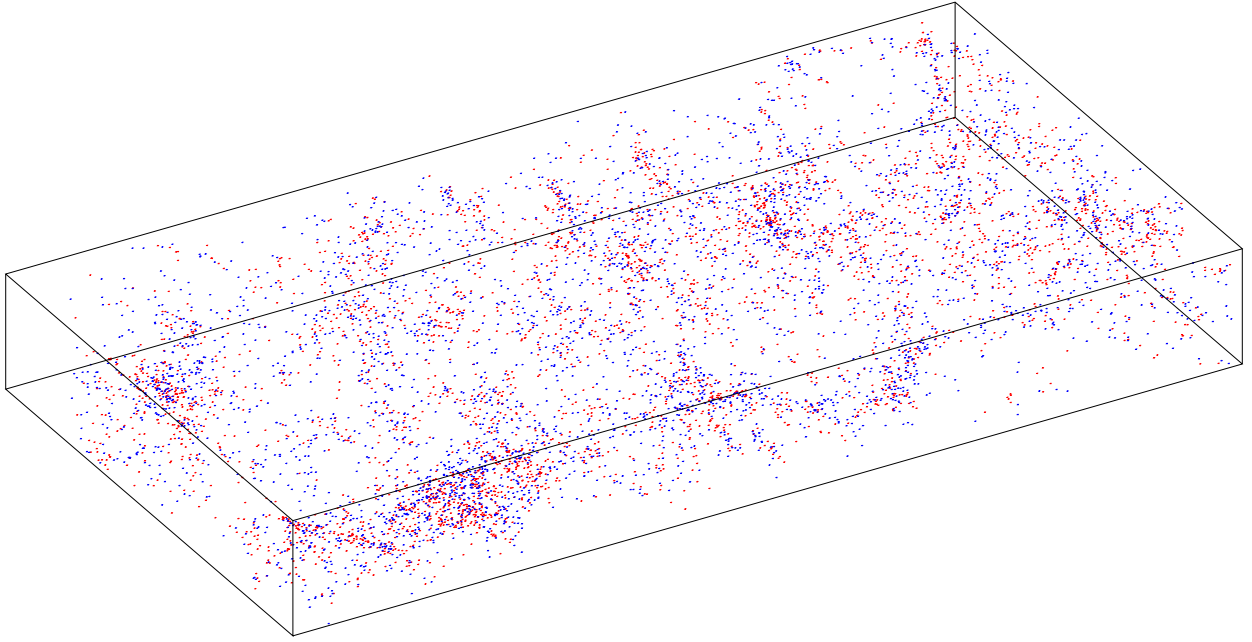}
\caption{Tridimensional plot of the galaxy sample used. 
Red dots correspond to early-type galaxies (population `1'), 
and blue dots to late-type galaxies (population `2'). The 
parallelepiped dimensions are $254\times133\times31$ \Mpc.}
\label{fig:slice}
\end{figure*}

We characterized the galaxies in the sample using the spectral 
classification parameter $\eta$ \citep{mad02a}. Lower 
values of $\eta$ correspond to
more passive or `early-type' galaxies, while larger values correspond 
to active or `late-type' ones. In order to avoid negative values of the marks, we defined
the mark used as $m = \eta + 10$. This shift does not affect our conclusions. 
Based on this $\eta$ parameter, we divided our sample in two populations,
following \citet{mad03a}: population `1' (passive galaxies) 
with $\eta \leq -1.4$, and population `2' (active galaxies) with 
$\eta > -1.4$. These subsamples contain $N_1 = 3828$  and  
$N_2 = 3913$ galaxies, respectively. We show the sample used in Fig.~\ref{fig:slice}.

In order to test the existence of mark segregation, we compared 
the results obtained for the different statistics
 with random relabelling simulations.
In these,  we keep the original positions of galaxies, but  redistribute
the marks randomly among them. This corresponds to a model in which 
clustering is independent of the mark,
or spectral type, of the galaxies.
We simulated $n=200$ realizations using the random relabelling method, 
and obtained their maximum and minimum values as function
of distance $r$ for each statistic. Deviations
of the observed statistics from this range of values correspond to 
rejection of the mark-independent clustering model at a pointwise significance of
$1 - \frac{2}{n+1} \simeq 99\%$ \citep{ili08a}.

\section{Clustering analysis methods}

Recently, the clustering dependence on luminosity, 
colour or morphology has been analysed by means
of the marked clustering statistics, that allow to study 
the galaxy clustering as a function of their properties, and
moreover this approach provides us with different measures of 
the correlation between the galaxy properties and
the environment \citep{ski08}.  The galaxy distribution 
is interpreted as a realisation of a marked point process
$X^M = \{ (\vec{x}_i, m_i)\}$, where the mark $m_i$ denotes an 
intrinsic property of the galaxy at position $\vec{x}_i$.
The mark can be the luminosity, the spectral type, the colour, etc. 
In general, today galaxy catalogues provide quantitative marks
ranging in a continuous interval rather than just a discrete 
characteristic like a galaxy being spiral or elliptical. 
In any case, we shall
also show how to use interesting second-order measures to 
disentangle clustering dependent characteristics of two
populations by dividing the sample in two parts using a significant 
value of the mark as threshold $m_{\rm thres}$ and
separating the two populations according to the value of the mark: 
population `1' with $m_i  \leq m_{\rm thres}$ and population `2'
with  $m_i  \geq m_{\rm thres}$.

We describe below the different methods we used to obtain information about
galaxy clustering segregation. They are the classical partial correlation functions
(for two discrete populations), the normalized mark correlation function and the mark
variogram (based on the use of continuous marks), and finally the mark connection function
(based on the use of discrete marks).

We computed the different statistics based on the estimation of
the second-order intensity function for the unmarked point 
process\footnote{Note that the relation between $\lambda_2(r)$ 
and the standard correlation function used in cosmology is 
$\xi(r)$ is $\lambda_2(r)=n^2[1+\xi(r)]$, where $n$ is the 
number density. The function $g(r)=1+\xi(r)$ is known as the pair correlation function
in spatial statistics.  We use the convention of denoting the estimators
by putting a hat $\hat{}$ on top of the symbol of a given function to distinguish
the estimator $\hat{\lambda}_{2}(r)$ from the theoretically defined function 
$\lambda_2 (r)$. Although this is not standard in cosmology, it is an extended
convention in spatial statistics, and it is quite useful when different 
estimators of a single function are discussed (see, e.g., \citet{pons1999}).} ($\lambda_2 (r)$)
 presented in \citet{sto94a}, \citet{pons1999}, 
and \citet{ili08a}, 
\begin{equation}
\hat{\lambda}_{2}(r) = \frac{1}{4 \pi r^2}
\sum_{i=1}^{N}    \sum\limits_{j=1\atop j\neq i}^{N}
 \frac{k(r-|\vec{x}_i-\vec{x}_j|)}
{V(W \cap W_{\vec{x}_i-\vec{x}_j})} \, ,
\label{eq:l2est}
\end{equation}
where $\vec{x}_i$ are the positions of the points,
 $k(\cdot)$ is a kernel function, and $V(W \cap W_{\vec{r}})$  
is the volume of the
window (the parallelepiped in our case) intersected with a 
version of itself shifted by the vector $\vec{r}$
(see Fig. 1 in \citealt{pons1999}).

In all our calculations we used the Epanechnikov kernel,
\[
k(x) = \left\lbrace\begin{array}{ll}
\frac{3}{4w} \left(1 - \frac{x^2}{w^2} \right) & \mbox{for}\, -w \leq x \leq w \\
0                                                                         & \mbox{otherwise}
\end{array} \right.
\]
with width $w=1$ \Mpc,  
and sampled the different functions with a step in $r$ of 0.5 \Mpc.
This compact kernel is very well suited for correlation analysis  \citep{pons1999}. We note, however, that the choice of a given kernel
is not crucial, while  the choice of the bandwith, $w$, is more important and plays the 
role of the binning
in the standard calculation of correlation functions, where a top-hat kernel is typically used as
default.

\subsection{Partial two-point correlation functions}

In the standard clustering analysis of the galaxy distribution, 
the two-point correlation function, $\xi(r)$, measures the clustering
in excess ($\xi(r)>0$) or in defect ($\xi(r)<0$)
relative to a Poisson distribution, for which $\xi(r)=0$. Whenever we want to compare
the clustering properties of different populations of galaxies encapsulated 
by their spatial correlations, we can consider the correlation function
restricted to a given population, which are called {\it partial} correlation 
functions. In fact, for two populations of
interest, one can consider three partial two-point correlation
functions, namely $\xi_{11}(r)$, $\xi_{22}(r)$, and $\xi_{12}(r)$. 
The first two are those mentioned above for types 1 and 2, 
while the cross-correlation function \citep{peeb80}
$\xi_{12}(r)$ measures the excess probability of finding a
neighbour of type `1' at distance $r$ from a given galaxy of 
type `2', or vice versa.

Based on equation~(\ref{eq:l2est}), the partial two-point correlation 
functions were estimated as
\begin{equation}
\hat{\xi}_{ij}(r) = \frac{1}{4 \pi r^2 \hat{n}_i \hat{n}_j}
\sum_{k=1}^{N_i}    \sum\limits_{l=1}^{N_j}
 \frac{k(r-|\vec{x}_k^{(i)}-\vec{x}_l^{(j)}|)}
{V(W \cap W_{\vec{x}_k^{(i)}-\vec{x}_l^{(j)}})} - 1\, ,
\label{eq:xiest}
\end{equation}
where $\vec{x}_k^{(i)}$ are the positions of galaxies of 
population $i$, and $\hat{n}_i = N_i /V$.

We estimated the error of the measured correlation functions using the
jackknife method \citep{norberg09}. 
We divided the data volume in 32 equal, nearly cubic,
sub-volumes. We generated the corresponding `mock' datasets omitting one of these sub-volumes
at a time, and calculated the correlation functions for these. The jackknife errors
for each scale, $\sigma_{ij}(r)$,  are then obtained as
\[
\sigma_{ij}^2(r) = \frac{N_k - 1}{N_k}
        \sum_{k=1}^{N_k} \left(\xi_{ij}^k(r) - \bar{ \xi}_{ij}(r)\right)^2 \, ,
\]
where $\xi_{ij}^k(r)$ is the partial correlation function $\xi_{ij}(r)$ of the `mock' dataset $k$, 
$\bar{ \xi}_{ij}(r)$ is the value averaged over these datasets, and $N_{k} = 32$.

\subsection{Normalized mark correlation function}

\Citet{sto94a}  introduced the normalized mark 
correlation function. To define this function let us first
define the quantity
\begin{equation}
\lambda_2^M [(\vec{x}_1,m_1),(\vec{x}_2,m_2)] dV_1 dm_1 dV_2 dm_2
\label{eq:l2def}
\end{equation}
as the joint probability that in the volume element $dV_1$ lies a
galaxy with the mark in the range $[m_1,m_1+dm_1]$ and that
another galaxy lies in $dV_2$ with the mark in $[m_2,m_2+dm_2]$ 
\citep{martsaar02}.
The normalized mark correlation function is
\begin{equation}
k_{mm}(r)=\frac{1}{\bar{m}^2\lambda_2(r)}\int \int m_1 m_2 \lambda_2^M
((\vec{x}_1,m_1),(\vec{x}_2,m_2)) dm_1 dm_2,
\label{eq:kmmdef}
\end{equation}
for $\lambda_2(r) \neq 0$, where $\bar{m}$ is the mean of the marks.

Despite its name the mark correlation function is not a strict
correlation function \citep{schlather01}, 
but it describes important
aspects of the spatial correlations of marks. 
A true mark correlation is a function given by Eq.~(\ref{eq:kmmdef}), but
replacing the product $m_1m_2$ by the product of the differences $(m_1-\bar{m})(m_2-\bar{m})$.
The normalizing denominator $\bar{m}^2$ must then be
replaced by $\sigma^2$, the variance of the marks. In any case,
$k_{mm}(r) < 1$
represents inhibition of the marks at the scale $r$. For example, in
forests it is typically found that trees with larger stem diameter
(mark) tend to be isolated, since they make use of much more ground
and sun-light resources than smaller trees.
Using luminosity as the mark, the opposite effect has been
found for the galaxy distribution, i.e., $k_{mm}(r)
>1$ at small scales \citep{beis00}, implying stronger
clustering of brighter galaxies at small separations, in agreement
with previous results showing this kind of segregation \citep{hamilton1988}.

We estimated the normalized
mark correlation function as
\begin{equation}
\hat{k}_{mm}(r) = \frac{1}{4 \pi r^2 {\bar{m}}^2 \hat{\lambda}_2(r) }
\sum_{i=1}^{N}    \sum\limits_{j=1\atop j\neq i}^{N}
 \frac{m_i m_j k(r-|\vec{x}_i-\vec{x}_j|)}
{V(W \cap W_{\vec{x}_i-\vec{x}_j})} \, .
\label{eq:kmmest}
\end{equation}

\subsection{Mark variogram}

The mark variogram, $\gamma(r)$ \citep{wal96a, beis00}, is a measure
of the similarity of the marks depending on the distance 
between galaxies. It is defined as 
\begin{equation}
\gamma(r)=\frac{1}{2 \lambda_2(r)}\int \int (m_1 - m_2)^2 \lambda_2^M 
((\vec{x}_1,m_1),(\vec{x}_2,m_2)) dm_1 dm_2 \, .
\label{eq:gammadef}
\end{equation}
When the clustering properties of a marked point pattern are independent 
of the marks, then the mark variogram $\gamma(r)$ is
constant and takes, naturally,  the value of the variance, $\sigma_m^2$, of the mark
distribution. In the presence of segregation, the fact that $\gamma(r) >
\sigma_m^2$ indicates that galaxy pairs at distance $r$ tend to
have different marks, while the contrary, $\gamma(r) < \sigma_m^2$, is an indication 
that these galaxy pairs tend to have similar marks.

We estimated the mark variogram as
\begin{equation}
\hat{\gamma}(r) = \frac{1}{8 \pi r^2 \hat{\lambda}_2(r) }
\sum_{i=1}^{N}    \sum\limits_{j=1\atop j\neq i}^{N}
 \frac{(m_i-m_j)^2 k(r-|\vec{x}_i-\vec{x}_j|)}
{V(W \cap W_{\vec{x}_i-\vec{x}_j})} \, .
\label{eq:gammaest}
\end{equation}

\subsection{Mark connection function}

A statistical tool to characterize the spatial correlation between
the marks of a point pattern with discrete marks is the mark
connection function $p_{ij}(r)$, which represents the conditional
probability to find two galaxies of type $i$ and $j$ at
positions separated by a distance $r$, 
under the condition that
at these positions there are indeed galaxies. This function yields
information different to that from the
partial correlation functions, $\xi_{ij}(r)$, 
as shown, for example, in \citet{ili08a}. By its definition it gives the 
relative frequencies of mark pairs
$(i,j)$ of distance $r$. While $\xi_{ij}(r)$ takes large values if
there are many $(i,j)$-pairs at distance $r$, $p_{ij}(r)$ is large if the
proportion of $(i,j)$-pairs in all pairs at distance $r$ is large. So
it may happen that for some $r$, $\xi_{ij}(r)$ has a minimum, but
$p_{ij}(r)$ has a maximum, if there is only a small number of point
pairs at distance $r$ in the whole pattern, but many of them are
exactly $(i,j)$-pairs. Experience shows that often $p_{ij}(r)$ is able to find finer structures
in point patterns than $\xi_{ij}(r)$, because of the
nature of $p_{ij}(r)$ as a conditional probability. 

If the marking is independent of clustering, 
then  $p_{ij}(r)$ are constant,
\begin{equation}
p_{ij}(r) = \left\lbrace \begin{array}{lcl}
2 p_ip_j & \mathrm{if} & i\neq j \\
p_i^2    &  \mathrm{if} &i=j\end{array}\right. \,.
\label{eq:random}
\end{equation}
Here $p_i$ is the probability that a randomly chosen galaxy is of
type $i$. The $p_i$ are estimated as
\[
\hat{p}_i = \frac{N_i}{N} \,.
\]
We calculated $p_{ij}(r)$ based on the estimation of the partial
correlation functions as
\[
\hat{p}_{ij}(r) = \hat{p}_i \hat{p}_j \frac{\hat{\xi}_{ij}(r) + 1}{\hat{\xi}(r) + 1} \, ,
\]
where $\hat{\xi(r)}$ is the two-point correlation function of the full sample.

\section{Results}

\subsection{Partial two-point correlation functions}

\begin{figure}[ht]
\centering
\resizebox{0.5\textwidth}{!}{\includegraphics*{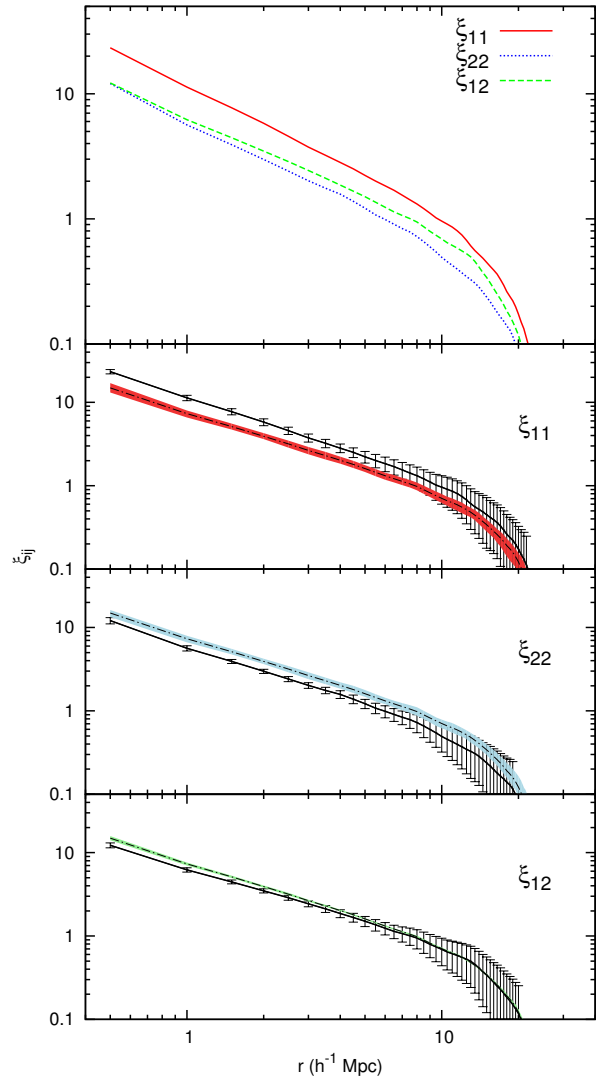}}
\caption{The partial two-point correlation functions, $\xi_{ij}(r)$,
estimated for population `1' (early-type) and population `2' 
(late-type) galaxies in our sample. The top panel shows the 
three functions together. The three
lower panels show each of them separately (solid lines
with error bars estimated using the jackknife method), 
together 
with a shaded band showing the
minimum and maximum values for the 200 realizations of the random 
relabelling simulation.
The dot-dashed lines correspond to $\xi(r)$ for the full sample, 
which is the expected value
of all $\xi_{ij}(r)$ in the absence of segregation.}
\label{fig:xiij}
\end{figure}

Fig.~\ref{fig:xiij} shows the three corresponding partial two-point correlation
functions, estimated according to Eq.~(\ref{eq:xiest}).
All three show clearly the high degree of clustering
within the pattern of galaxies. It is obvious that the correlation
function for the type `1' passive galaxies is steeper than for the type `2'
active galaxies as well as for the (1,2) pairs. This result corroborates
the spectral segregation detected by \citet{mad03a} for the 2dFGRS.

\subsection{The normalized mark correlation function}

\begin{figure}[ht]
\centering
\resizebox{0.5\textwidth}{!}{\includegraphics*{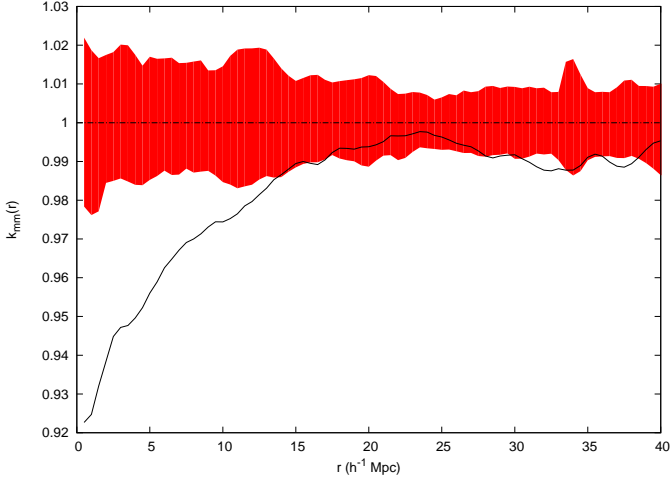}}
\caption{Normalized mark correlation function, $k_{mm}(r)$, 
for our sample (solid line).
The shaded band shows the minimum and maximum values for the 
200 realizations of the random relabelling
simulation, while the dot-dashed line correspond to the value 
for the case with no segregation, $k_{mm}(r) = 1$.}
\label{fig:kmm}
\end{figure}

The $k_{mm}(r)$ for our sample, estimated according to
Eq.~(\ref{eq:kmmest}), is shown
in Fig.~\ref{fig:kmm}.
The curve for $k_{mm}(r)$ shows a weak negative correlation
or spatial inhibition: $k_{mm}(r) < 1$. The
range of correlation is about 20 \Mpc, where $k_{mm}(r)$ 
gets values close to
1. It is interesting to compare this result with the $k_{mm}(r)$-function 
shown in \citet{beis00} using the galaxy absolute luminosity  
$L$ as the mark.
They obtain an increasing behaviour of $k_{mm}(r)$ at small 
scales with
$k_{mm}(r) >  1$ for $r <  12 \,h^{-1}$ Mpc, showing that bright galaxies
are stronger correlated than faint ones.  In our case, the tendency of the values
of $k_{mm}(r)$ to be smaller than 1 at short scales indicates that the price
galaxies have to pay for being close together is to have reduced values of the
marks, i.e., being more passive.

\subsection{The mark variogram}

\begin{figure}[ht]
\centering
\resizebox{0.5\textwidth}{!}{\includegraphics*{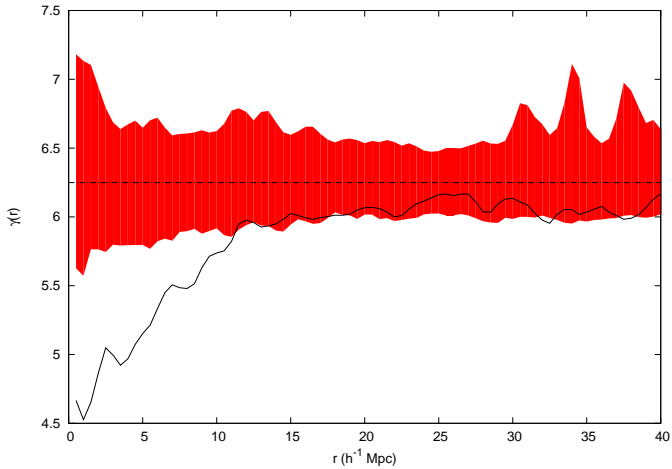}}
\caption{Mark variogram, $\gamma(r)$, for our sample (solid line).
The shaded band shows the minimum and maximum values for the 
200 realizations of the random relabelling
simulation, while the dot-dashed line corresponds to the value 
for the case with no segregation, $\gamma(r) = \sigma_m^2 = 6.25$.}
\label{fig:gamma}
\end{figure}

In Fig.~\ref{fig:gamma}, we show the mark variogram for our sample,
obtained according to Eq.~(\ref{eq:gammaest}).  This
function is monotonously increasing. In this case the interpretation is
straightforward:  $\gamma(r)$ shows that,
for separations $r \lesssim 10$ \Mpc, galaxy pairs tend to have similar marks, that is,
similar spectral type. 

This result is partially explained with the previous one shown
by the $k_{mm}$ function:  galaxies close together exhibit
smaller values of the attached mark (spectral type).

\subsection{The mark connection function}

\begin{figure}[ht]
\centering
\resizebox{0.5\textwidth}{!}{\includegraphics*{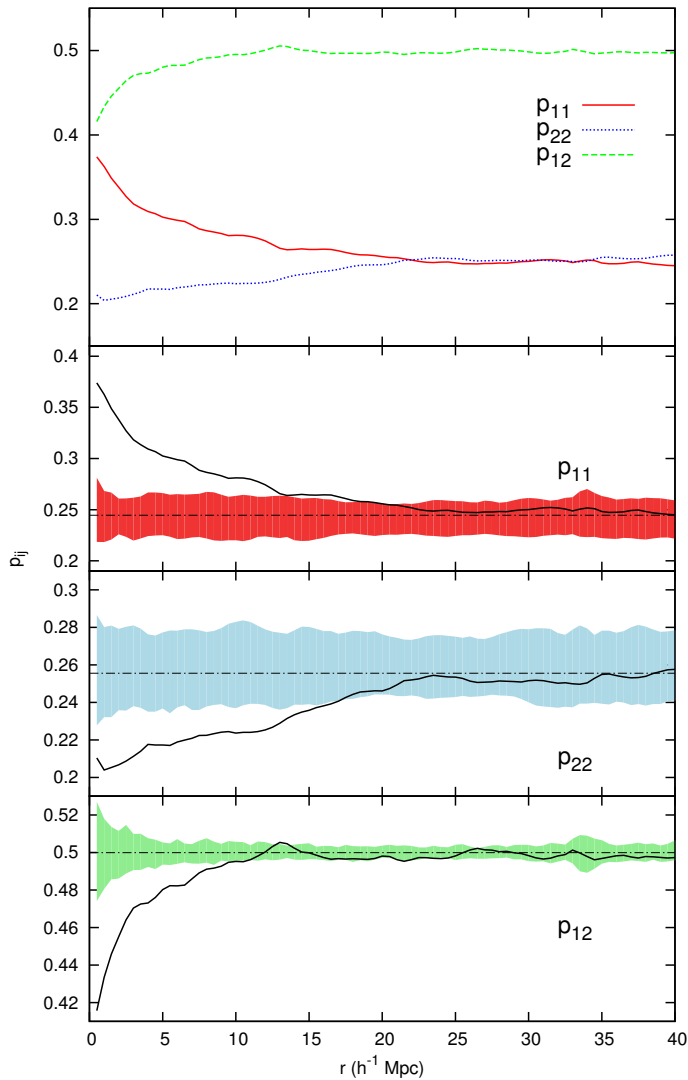}}
\caption{Mark connection functions, $p_{ij}(r)$ obtained for 
`early-type' (population `1') and `late-type' (population `2') 
galaxies in our sample. The top panel shows the three functions 
together. The three bottom panels show $p_{11}(r)$, $p_{22}(r)$, 
and $p_{12}(r)$ separately (solid lines), together with the shaded 
band showing the minimum and maximum values for the 200 realizations 
of the random relabelling simulation. The dot-dashed lines correspond 
to the expected values for the random labels case,
according to Eq.~\ref{eq:random}.
}
\label{fig:pij}
\end{figure}

We show the $p_{ij}(r)$ obtained for the 2dFGRS galaxies, 
together with the results of our random relabelling 
simulations, in Fig.~\ref{fig:pij}. The first panel shows 
very neatly that, for scales $r \lesssim 20$ \Mpc, the clustering 
of early-type galaxies is stronger than the clustering of late-type 
galaxies. 
The three bottom panels show that the deviation
of the observed $p_{ij}(r)$ from the case of random labelling 
is significant at these scales.

Moreover, the figure shows clear differences in the spatial
correlations of galaxies of the two types. In an overall clustering
of all galaxies, we can outline that:

\begin{enumerate}
\item
Galaxies of type `1' (passive or early-type) are strongly 
clustered up to distances of 20 \Mpc.

\item
The conditional probability to find two galaxies of type `2' 
(active or late-type) at two positions separated by a distance $r$ 
(under the condition that at these locations are galaxies) is smaller 
than the same probability for random labelling of the marks 
for scales $r \lesssim 20$ \Mpc.
\item
Galaxy pairs having one member of type `1' and the other member of type `2' 
are less frequent than for random labelling up to distances of 10 \Mpc.
\end{enumerate}

In summary, all galaxies form a highly clustered pattern. In
this pattern, the passive galaxies tend to be close to other passive
galaxies, while positioning of active galaxies is less affected by other
active galaxies. However, they tend
to avoid positions close to passive galaxies.

This shows clearly the power of the mark connection function as an
analytical tool in comparison to the partial pair correlation
function. While, for the untrained eye, the curves in Fig. 2 are
quite similar and show little structure, the curves in Fig. 5 give
valuable information about the inner structure of the mark
distribution. Obviously, the idea to consider characteristics of the
nature of conditional probabilities helps to divulge structural
details which would be otherwise overlooked.

The problem are the mutual positions, given the positions of all
galaxies without mark information. Since the three partial two-point
correlation functions, shown in Fig. 2, are different for a large range of scales, 
the marking with marks 1 and 2
can not be an independent marking, where every galaxy obtains its
mark randomly, independent of the other galaxies. 
In contrast, there
must exist a spatial correlation between the marks. As it has been 
shown in Fig. 5,  the mark connection function is the appropriate tool
to measure this correlation.

\section{Conclusions}

We have used a volume-limited galaxy sample from the 2dFGRS to test different
statistical measures used to disentangle mark segregation in the distribution
of the galaxies. The mark attached to each galaxy of the sample was
its spectral type $\eta$. For some of the statistics,  the value of the
mark enters directly into the functions used for measuring segregation:
the normalized mark correlation function $k_{mm}(r)$ and the mark variogram
$\gamma(r)$.  For other functions, as the  partial correlation functions
or the mark connection function, the sample has been split into two populations
corresponding to passive or early-type galaxies with $\eta \leq -1.4$ 
and active or late-type galaxies
with    $\eta>-1.4$. We summarise our results as follows:
\begin{enumerate}
\item
The partial correlation functions, including the cross-correlation function, inform us
about the degree of clustering of each population separately. It shows that passive galaxies
exhibit stronger clustering at small separation. Nevertheless, there is no information about
the spatial correlation between the marks.
\item
The normalized mark correlation function shows that having
smaller values of the marks, i.e., smaller values of spectral type
(being more passive), is a clear condition for galaxies being
close to each other in the overall clustering pattern.
\item
The mark variogram, in addition, shows that at small separations galaxy pairs tend to have
similar marks.
\item
The mark connection function has been introduced here for the first
time in the analysis of the marked galaxy distribution.  The
function $p_{ij}(r)$ measures the conditional probability to find at
two positions, separated by a distance $r$, a galaxy of type `i' and
a galaxy of type `j' under the condition that at these positions there
are
indeed galaxies. This function yields information different from
that of the partial correlation functions $\xi_{ij}(r)$. This more
sophisticated measure, having a nature of conditional quantities, is
an efficient statistical tool to characterize the spatial
correlation between the marks,  filtering out the relative
frequencies of the mark pairs $(i,j)$ at distance $r$.

Applied on the 2dFGRS volume-limited sample, the mark connection function
clearly shows that passive galaxies are clustered up to distances of
20 \Mpc, while active galaxies exhibit weak spatial anticorrelation of the mark
up to distances of 20 \Mpc. Mixed pairs are less frequent up to distances of 10 \Mpc.
\end{enumerate}

\begin{acknowledgements}
First, we thank the anonymous referee for detailed and constructive criticism and
suggestions.
We are pleased to thank the 2dFGRS Team for the publicly available
data releases.  We thank D. Croton for the 2dFGRS samples and the mask data and
M.~J. Pons-Border\'{\i}a for comments and suggestions.
This work has been supported by Spanish Ministerio de Ciencia e Innovaci\'on
CONSOLIDER projects
AYA2006-14056 and CSD2007-00060, including
FEDER contributions, and by the Generalitat 
Valenciana  project of excellence PROMETEO/2009/064.
PAM acknowledges support from the
Spanish Ministerio de Educaci\'on through a FPU contract.

\end{acknowledgements}

\bibliographystyle{aa}
\bibliography{mark}
\end{document}